\documentclass[sigconf]{acmart}
\usepackage{graphicx}
\usepackage{hyperref}

\usepackage{multirow}
\usepackage{xspace}
\usepackage{mdframed}
\usepackage{csquotes}
\usepackage{subcaption}
\usepackage{csquotes}

\usepackage{pythonhighlight}
\usepackage{booktabs}
\usepackage{xcolor}
\usepackage{enumitem}

\author{Owen Tucker}
\affiliation{%
  \department{Department of Computer Science}
  \institution{Emory University}
  \city{Atlanta}
  \country{USA}
}
\email{owen.tucker@emory.edu}

\author{Lily Wang}
\affiliation{%
  \department{Department of Computer Science}
  \institution{Emory University}
  \city{Atlanta}
  \country{USA}
}
\email{lily.wang2@emory.edu}

\author{Ruochen Kong}
\orcid{0009-0006-0329-8019}
\affiliation{%
  \department{Department of Computer Science}
  \institution{Emory University}
  \city{Atlanta}
  \country{USA}
}
\email{ruochen.kong@emory.edu}

\author{Haley Stone}
\affiliation{%
  \institution{MRC University of Glasgow Centre for Virus Research}
  \city{Glasgow}
  \country{United Kingdom}
}
\affiliation{%
  \department{School of Computer Science and Engineering}
  \institution{University of New South Wales}
  \city{Sydney}
  \country{Australia}
}
\email{haley.stone@glasgow.ac.uk}

\author{Jing Du}
\affiliation{
  \department{School of Computer Science and Engineering}
  \institution{University of New South Wales}
  \city{Sydney}
  \country{Australia}
}
\email{jing.du2@unsw.edu.au}

\author{Matthew Scotch}
\affiliation{%
  \department{College of Health Solutions}
  \institution{Arizona State University}
  \city{Tempe}
  \country{USA}
}
\email{matthew.scotch@asu.edu}

\author{Flora D. Salim}
\affiliation{%
  \department{School of Computer Science and Engineering}
  \institution{University of New South Wales}
  \city{Sydney}
  \country{Australia}
}
\email{flora.salim@unsw.edu.au}

\author{Taylor Anderson}
\affiliation{%
  \department{Department of Geography and Geoinformation Science}
  \institution{George Mason University}
  \city{Fairfax}
  \country{USA}
}
\email{tander6@gmu.edu}

\author{Chandini Raina MacIntyre}
\affiliation{%
  \department{The Kirby Institute \\ Faculty of Medicine \& Health}
  \institution{University of New South Wales}
  \city{Sydney}
  \country{Australia}
}
\email{r.macintyre@unsw.edu.au}

\author{Andreas Z\"ufle}
\affiliation{%
  \department{Department of Computer Science}
  \institution{Emory University}
  \city{Atlanta}
  \country{USA}
}
\email{azufle@emory.edu}
\usepackage{amsmath}
\usepackage{float}

\copyrightyear{2026}
\acmYear{2026}
\setcopyright{cc}
\setcctype{by}
\acmConference[SIGSPATIAL '26]{The 34th ACM International Conference on Advances in Geographic Information Systems}{November 03--06, 2026}{Riverside, CA, USA}
\acmBooktitle{The 34th ACM International Conference on Advances in Geographic Information Systems (SIGSPATIAL '26), November 03--06, 2026, Riverside, CA, USA}
\acmDOI{10.1145/3841645.3843291}
\acmISBN{979-8-4007-2950-8/2026/11}

\begin{document}




\title{Inferring Locations and Correcting Bias in Infectious Disease Data Using Human Mobility}

\begin{abstract}

Understanding and reconstructing infectious disease transmission remains challenging due to sparsely sampled and biased epidemiological case data: Only a small fraction of infectious disease cases are reported to health authorities and an even smaller fraction of cases are sent to labs for genomic sequencing. 
In this work, we study the problem of inferring unobserved transmission events by integrating pathogen genomic data with human mobility information. Intuitively, even if we don't have case data for a certain region such as a census block group, we can use massive-scale individual-mobility data to understand where people may be exposed and infer where unobserved cases may be located. 
Assuming only a subset of cases are reported and sequenced, we reconstruct a phylogenetic tree from observed cases and transform it into a partial transmission tree using the \texttt{outbreaker2} phylogenetic inference tool. This tree contains inferred transmissions based on the genetic distance of the observed cases. As these inferred nodes have unknown locations, we then propose a mobility-aware belief propagation approach to estimate their locations using human mobility data. 
Experiments on simulated and real-world genomic surveillance data that include biased sampling demonstrate that incorporating mobility data improves reconstruction of inferred transmission trees; debiasing from the observed distribution by reducing cross-entropy by $39.4\%$ and $16.1\%$ respectively. Results on real-world COVID-19 sequence data show promise of combining genomic and mobility data for more reliable location inference, with potential implications for outbreak analysis and public health decision-making.

\vspace{-.45cm}   
\end{abstract}

\ccsdesc[500]{Applied computing~Health informatics}

\keywords{{Epidemiology, Infectious Diseases, Phylogeography, Mitigating Bias, AI for Public Health}}

\maketitle

\section{Introduction}
\label{sec:introduction}


Pathogen genomic surveillance is now widely used in outbreak investigation to identify related infections, monitor viral evolution, and reconstruct patterns of disease spread~\cite{armstrong2019pathogen,bedford2020cryptic,fauver2020coast}. When linked with sampling dates and locations, pathogen sequences can provide insight into how observed infections are connected across time and space, supporting phylogenetic analyses that identify related cases and reconstruct evolutionary relationships among observed infections~\cite{worby2017shared,boussau2010genomes}. However, genomic surveillance data are not a complete record of transmission. They reflect who was tested, which samples were sequenced, and where cases were recorded, all of which may be shaped by public health capacity, healthcare access, sequencing priorities, and reporting practices. As a result, the spatial distribution of sequenced cases may differ substantially from the true distribution of infections in an outbreak.

However, phylogenetic analyses are only as representative as the surveillance data on which they are based. When sequencing is sparse or uneven across populations and locations, the resulting phylogeny may provide a clear evolutionary picture of sampled cases while still giving an incomplete or biased view of the underlying outbreak. In addition, while phylogenetic trees capture general evolutionary relationships, they do not directly encode person-to-person transmission pathways \cite{colijn2014phylogenetic}. Moreover, sampling bias can distort our understanding of outbreak dynamics. For example, differential testing or sequencing across populations can make some groups or regions appear over- or under-represented in surveillance data, regardless of their true infection burden~\cite{griffith2020collider}. Individuals and locations that appear in phylogenetic trees represent only a small subset of infections: those that were detected, tested, selected for sequencing, successfully sequenced, and linked to usable metadata. These biases also have direct consequences for public health messaging. If messages are based only on reported or sequenced cases, public health agencies may overemphasize areas with greater testing or sequencing capacity while failing to reach communities where transmission is ongoing but under-detected. Mobility-informed reconstruction can help distinguish reported disease burden from inferred transmission risk, supporting more equitable and geographically precise risk communication.

One way to narrow this gap is to use phylogenetic information to reconstruct plausible transmission~\cite{didelot2014bayesian}. These methods translate genetic and temporal relationships among sampled cases into hypotheses about who infected whom and when transmission may have occurred. Given a reconstructed transmission tree, we can then ask where unobserved intermediate infections were most likely to have occurred, which can aid in public health agencies in identifying under-detected transmission. 
However, transmission-tree reconstruction is challenging because real-world outbreak data are sparse and selectively observed:
\begin{enumerate}
\item A significant portion of cases of an infectious disease may be entirely asymptomatic. 
For example, this fraction was reported to be ~40\% for SARS-CoV-2~\cite{ma2021global}, 
\item Early in the SARS-CoV-2 pandemic in China, only about 14\% of infections were estimated to have been reported~\cite{li2020substantial}.
\item Among reported cases, only a small fraction is sequenced. For SARS-CoV-2, the sequencing rate was typically around 0.5\% (one in two hundred) across most countries worldwide~\cite{brito2022global}.
\end{enumerate}
These layers of under-ascertainment introduce substantial uncertainty into transmission inference and make it difficult to identify missing infections and transmission links. For example, two genetically related cases observed in different cities may imply unobserved intermediate infections, but the number, timing, and locations of those infections may remain unknown. Moreover, infectious disease spread is inherently shaped by human mobility. Daily movement patterns, such as commuting, travel, and social interactions across space, strongly influence where and how transmission occurs. Mobility data derived from transportation records, mobile devices, or census statistics have been widely used in population-level epidemiological models to forecast disease spread and evaluate intervention strategies. Despite its demonstrated relevance, mobility information is rarely integrated into phylogeny-based transmission inference, where most existing methods rely primarily on genetic data and, in some cases, limited temporal or contact information. As a result, currently inferred transmission trees may be genetically consistent yet spatially implausible. This observation motivates the need for inference frameworks that jointly consider genomic similarity and spatial proximity. Incorporating mobility data provides an additional, epidemiologically meaningful signal that can constrain transmission hypotheses and reduce uncertainty when genetic evidence alone is weak.

\begin{figure*}
    \centering
    \includegraphics[width=.9\linewidth]{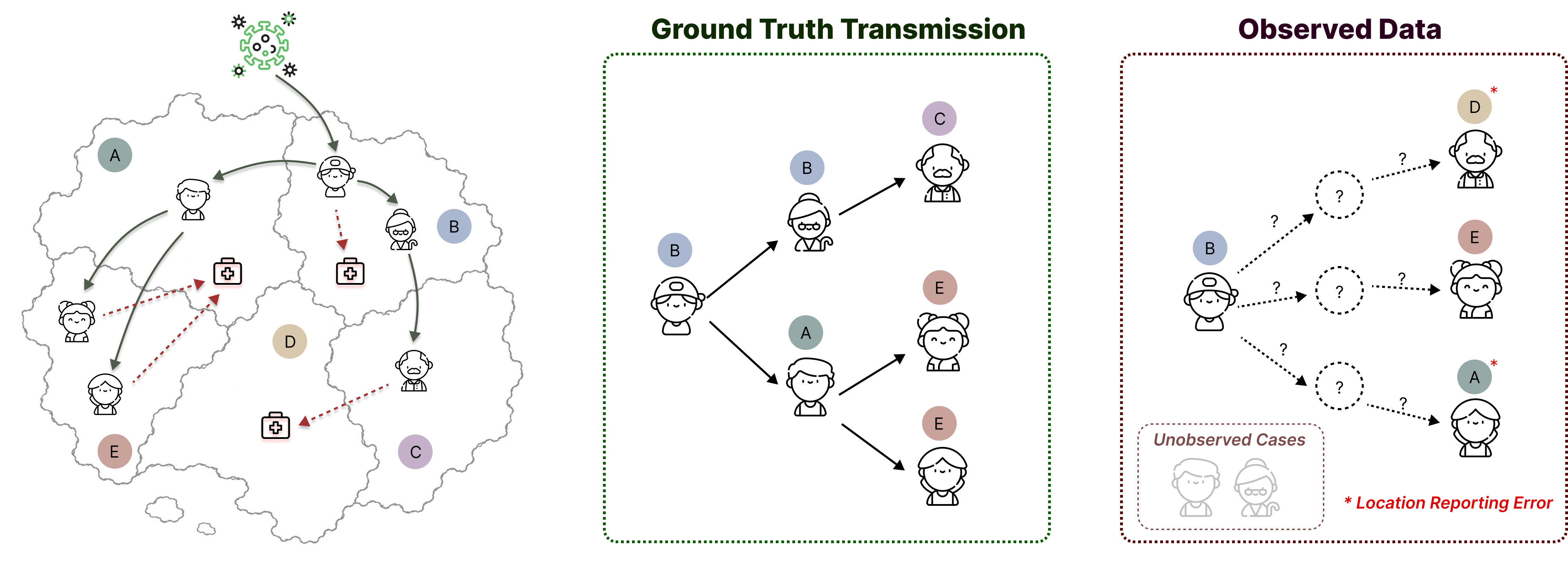}
    \caption{Problem Definition: \textbf{Left:} Inferring the spatial distribution of cases using sparsely sampled and biased data: An infectious disease spreads across individuals. Some of these cases are reported (red dashed lines) at laboratories which may be in different spatial regions. \textbf{Middle:} Ground truth transmission tree including all cases and their locations (A-E). \textbf{Right:} Observed cases with missing data (two cases are unobserved) and location bias (two cases are observed in incorrect locations).}
    \label{fig:prob}
\end{figure*}

In this work, we combine epidemiological transmission-tree reconstruction with human mobility data to estimate the likely locations of unobserved infections using belief propagation. For each unobserved infection, the model estimates a posterior distribution over candidate locations and assigns a maximum a-posteriori (MAP) location based on this marginal distribution. Unlike phylogeographic tools such as BEAST~\cite{bouckaert2014beast}, which model geographic location as a discrete trait evolving along a phylogeny, our approach uses human mobility flows as transition probabilities over epidemiologically plausible transmission pathways, enabling geographic reconstruction of potential unobserved infections. By incorporating mobility flows, the model aims to identify where unobserved infections are epidemiologically plausible given observed transmission relationships and population movement and may complement traditional contact tracing and genomic outbreak investigation~\cite{wang2022effectiveness}. Specifically, this paper makes the following contributions:
\vspace{-0.1cm}
\begin{itemize}
    \item We propose a mobility-informed phylogeographic pipeline for early-outbreak geographic characterization. We propose a pipeline that combines income-biased genomic surveillance data alongside standard epidemiological inference tools to produce reconstructed transmission trees in early-stage outbreaks. We extend this by incorporating human mobility data to infer the location of missing cases. We make our code available at: \url{https://github.com/OwenTucker/Phylogeographic-Inference-Model}
    \item Through evaluating our method on both simulated and real-world datasets, we demonstrate that mobility-informed reconstruction achieves lower cross-entropy against true case distribution when compared to the observed case distributions, with an average reduction of $39.4\%$ on simulated data, and $16.1\%$ on real-world data. 
    \item Lastly, we present an ablation study investigating how initialization of priors impacts reconstruction quality among simulated trees
\end{itemize}

In the following sections, we begin by illustrating our problem definition in Section~\ref{sec:problem}, we then briefly summarize the existing studies in phylogenetic location inference in Section~\ref{sec:related_works}. We explain the details of our approach in Section~\ref{sec:methodology}. Section~\ref{sec:data} describes the datasets we used to evaluate our method, and Section~\ref{sec:experiment} provides the experimental setup as well as our experiment results, and we conclude in Section~\ref{sec:conclusion}.

\section{Problem Definition}
\label{sec:problem}

In this section, we define our research problem in detail first using an illustrative example and then using a formal problem definition. Figure~\ref{fig:prob} (left) shows a toy example of how real-world data are derived from the underlying transmission process and how these observed data may differ from the ground truth using five regions and six individuals. As shown in this figure, an infectious disease spreads among the five regions (A, B, C, D, and E) and six individuals are infected, namely a middle-aged man in Region A, a boy and an old lady in Region B, an old man in Region C, and a girl and a middle-aged lady in Region E. Among these six individuals, only four (the boy, the old man, the girl, and the middle-aged lady) reported their cases and had their cases sequenced at a laboratory. With full knowledge of the disease transmission dynamics, we should have a ground-truth transmission tree shown in Figure~\ref{fig:prob} (middle). However, the data we observe is not perfect: In this example, two of the individuals did not report, meaning they were not known to be infected. Among the ones who reported, the location of cases of the old man and the middle-aged lady are incorrectly reported in the region of their laboratory instead of their home locations. Such errors leading to data bias, also known as collider biases~\cite{griffith2020collider}, exist significantly in real-world datasets. Consequently, we never have the full knowledge of infectious disease spread in the real-world. Motivated by the real-world scenario discussed above, we break the problem down into two steps, 1) inferring the locations of known but unreported cases, and 2) reconstructing the full transmission dynamics while accounting for unknown and unobserved cases. 


Formally, we define a case as a tuple, $c_i=(l_i,t_i,s_i)$, where $l_i,t_i$and $s_i$ denote the location, infection time, and genomic sequence of case $i$, respectively. Let $D_{obs}$ represent the collection of all fully observed and sequenced cases; 
and let $D_{GT}$ represent the underlying ground truth of all infectious disease cases, including those that were not observed. Our goal is to infer unknown cases and their locations in $D_{GT}$ given the observed in $D_{obs}$, formally:

\begin{definition}[Problem 1: Estimation of Unobserved Cases]
 Given sequenced cases with known locations $D_{obs}$, and cases with inferred location $D_{seq}$, our goal is to estimate the locations of unobserved cases, formally:
 $$\{(l_i,t_i,s_i)\}_{i \in D_{obs}} \rightarrow \{(\hat{l_j},\hat{t_j})\}_{j \in D_{GT}}$$
\end{definition}
We emphasize that our goal is not to infer the genetic sequence of the unobserved cases, which is the focus of the field of Phylogenetics~\cite{turakhia2022pandemic,perera2021reconstructing}. Rather, our goal is to estimate only the location of missing cases, which is the focus of the field of Phylogeography~\cite{rito2020phylogeography,sallard2021tracing}. Our main contribution to this field is leveraging massive-scale human mobility data as a prior to inform our model where people go and where transmission may be more or less likely to occur.

\section{Background and Related Work}
\label{sec:related_works}
Existing approaches relevant to this study span three areas: genomic reconstruction of transmission trees, phylogeographic inference of pathogen movement, and mobility-informed spatial modelling of infectious disease spread. We review these areas before describing the probabilistic tools used in our framework.

In this section, we first define phylogenetic trees, and how they differ from transmission trees. We then review methods that relate phylogenetic trees to transmission trees, including \texttt{outbreaker2}---a statistical model that uses sequence data to capture transition dynamics of densely sampled outbreaks. Finally, we discuss continuous-time Markov chains as our model for spatial transitions, and belief propagation as the inference algorithm we use for spatial inference on transmission trees.

\subsection{Phylogenetic Trees}
A key aspect of epidemiological modeling within infectious disease outbreaks is the ability to trace the spread of a pathogen across a population. The goal is to create a transmission tree, that connects individual cases by determining who infected who. This is often modeled temporally, where outbreaks are depicted in a tree-like structure and branch lengths represent time between observed cases. Traditionally, these methods have relied on the tedious task of contact tracing \citep{hall2016using}, and with new attempts at digital contact tracing proving to be an uncertain method \cite{cochrane1996digital}, recent methods have focused on utilizing the newly available genomic data made available by COVID-19. 


\begin{figure}
    \centering
\includegraphics[width=0.5\textwidth]{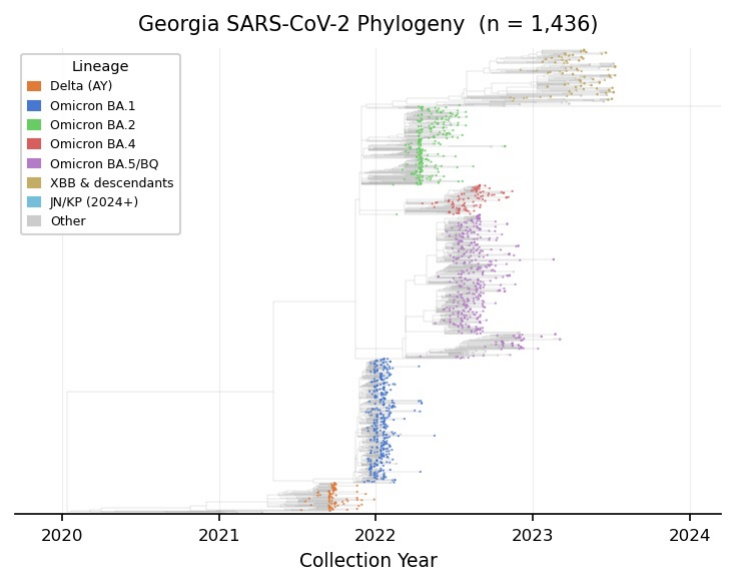}\caption{Real-world Georgia COVID data\citep{khare2021gisaid}}
    \label{fig:phylotree}
\end{figure}

As outbreaks spread, pathogens accumulate mutations. These result in changes to the genetic sequence of a virus. The accumulation of mutations results in genetically distinct cases. This can occur through high rates of mutation, and gradually through time. When analyzing two sequences from a pathogen, the proximity of their relationship can be estimated through comparing the amount of mutations. This principle is the foundation for phylogenetic trees, where sequenced cases are connected through genetic similarity. Reconstructing transmission trees from phylogenetic trees requires connecting sequence evolution and person-to-person transmission. Methods differ in assumptions about within-host diversity and how representative a sampled sequence is of the pathogen population infecting a given host \cite{hall2016using}. 

Figure~\ref{fig:phylotree} illustrates a phylogenetic tree generated using genomic data from GISAID (detailed in Section~\ref{sec:gisaid}) for the state of Georgia, USA. In this visualization, the x-axis represents time, while the y-axis carries no biological meaning other than to structure the tree topology, which groups cases sharing the same genetic lineage. The leaf nodes denote observed clinical cases. Crucially, a phylogenetic tree reflects genetic similarity rather than direct epidemiological transmission; thus, two leaf nodes sharing a parent node were not necessarily infected by the same source. 

Furthermore, a phylogenetic tree captures only a minute fraction of the true disease burden. Prior epidemiological modeling indicates that only an estimated $9.2\%$~\cite{anand2020prevalence} of actual cases are formally reported~\cite{noh2021estimation}. Among these reported cases only a small subset of approximately $1\%$ undergo genomic sequencing~\cite{brito2022global}. Consequently, a phylogenetic tree observes roughly $0.092\%$ (fewer than one in a thousand) of all true cases. This severe data sparsity motivates our core research question: Can we infer the spatial locations of the remaining $99.9\%$ of unobserved cases to improve infectious disease predictions and effectively guide public health interventions?


\subsection{Inferring Transmission Trees}
Relating phylogenetic trees to transmission trees is a well-established, yet challenging problem \citep{ypma2013relating, klinkenberg2017simultaneous}. The difficulty lies in the inherent differences between the two---phylogenetic trees capture the evolutionary development between cases, while transmission trees provide case-to-case transmission events between hosts. Most methods take advantage of the availability of whole-genome sequencing as samples become more and more available during an outbreak, and probabilistically construct transmission links through the timing and genomic information available. Furthermore, recent methods incorporate supporting proximal information, such as contact tracing data \citep{klinkenberg2017simultaneous}. 

\texttt{outbreaker2}~\citep{jombart2014bayesian} is a probabilistic model that reconstructs who-infected-whom by embedding pathogen genetic similarity within a Bayesian data-augmentation model that jointly infers the infection time and transmission links of estimated intermediate cases. Rather than treating the phylogeny itself as the transmission tree, outbreaker2 uses pairwise genetic distances (or sequence likelihoods) together with sampling dates and optional contact information to define a probabilistic transmission likelihood. The model couples three components: (i) a temporal prior derived from the generation-time distribution that constrains which infectors are epistemologically plausible; (ii) a mutation model that relates observed sequence differences to the time elapsed along transmission; and (iii) an observation process that allows for missing or unsampled intermediates. Inference is performed via Markov chain Monte Carlo, which iteratively proposes modifications to ancestors, infection dates, and importation status, thereby sampling from the posterior distribution over transmission trees. This formulation allows outbreaker2 to quantify uncertainty, integrate heterogeneous data sources, and scale to outbreaks where genomic similarity alone would be insufficient to uniquely resolve person-to-person spread. 

However, phylogenetic reconstruction from sampled sequences inherently yields incomplete transmission networks. Although \texttt{outbreaker2} was designed under the assumption of densely sampled outbreaks, we use it purely as a tree reconstruction tool, as it infers transmissions and infection times from genomic and temporal data but does not itself attempt to infer the spatial location of unsampled cases. While sparse sampling degrades the precision of inferred transmissions and infection time estimates, \texttt{outbreaker2} does not require complete sampling to produce a usable tree topology. We therefore treat the reconstructed tree and the inferred intermediate cases as a starting point, and use mobility-informed belief propagation to recover the spatial information that \texttt{outbreaker2}'s model does not provide.

\subsection{Belief Propagation}

Sum-Product Belief Propagation \citep{pearl2022reverend} is a messaging algorithm for inference on graphical models, specifically on Bayesian networks. The algorithm operates between two phases i) an upward pass where messages propagate from leaves towards the root and ii) a downward pass where messages propagate from root to leaves. 
The algorithm produces exact marginals for trees, making it well-suited for location-inference for nodes that have been generated in the transmission tree inference step and for which their location is unknown. An important aspect of initialization is setting priors for both known and unknown nodes. We fix sampled cases to their known origin, and experiment with multiple priors on unsampled cases generated from \texttt{outbreaker2}. After convergence, the marginal belief at each node is proportional to the product of all incoming messages. 

\subsection{Continuous-Time Markov Chains}
\label{subsec:ctmc}
A continuous-time Markov chain (CTMC) models instantaneous transitions between states that can occur at any continuous time. A CTMC is characterized through a rate matrix $\mathbf{Q}$ where $Q_{i,j} \ge 0, i\neq j$ represents the instantaneous rate of transition from state $i$ to state $j$, and the  diagonal value of rate matrix $\mathbf{Q}$ is the negative sum of its row, or $Q_{ii} = - \sum_{j\neq i} Q_{ij}$. 

The probability of transition from state $i$ to state $j$ over a given interval of time $\tau$ is given by the matrix exponential:

\begin{equation}
P_{ij}(\tau) = [\mathbf{P}(\tau)]_{ij} = [e^{\mathbf{Q}\tau}]_{ij}
\end{equation}

In our work, we round each branch length to four decimal places and cache $\text{expm}(\mathbf{Q\tau})$ to reuse for branches with near identical lengths. This transition matrix $\mathbf{P}(\tau)$ serves as the message passing factor between nodes during belief propagation, where $\tau$ corresponds to the branch length separating two nodes in the transmission tree inferred by \texttt{outbreaker2}.

\section{Datasets}
\label{sec:data}
In this section, we describe the sources and processing of the human mobility data used to build our model, as well as the real-world and simulated datasets used to evaluate our approach.

\subsection{Human Mobility Flow}
\label{sec:mobility}
We processed and aggregated mobility data from Advan~\cite{advan}, which tracks trajectories from millions of mobile phone users and reports daily visitor counts between Census Block Groups (CBGs), to construct aggregated human mobility flows at the county level. In our model, we assume a disease transmission is occurring when travelers from different regions co-locate within the same area, for example having lunch in the same restaurant. In other words, we assume that infection events are not restricted to individuals' home regions. An infected individual may subsequently travel to different regions on later days, enabling further spread. To capture this process, we calculate the destination distribution of individuals traveling from a given county to others, as well as the source distribution of individuals arriving in a county from elsewhere. Then the inter-county spread probability is derived by multiplying the destination and source distributions. Conceptually, the probability of disease transmission from county $A$ to $B$ is equivalent to the chance that the residents of both $A$ and $B$ visit the same intermediate location $X$. Let $M$ denote the mobility flow matrix, where $M_{ij}$ represents the flow from county $i$ to county $j$. We compute the row-normalized and column-normalized matrices as
\begin{align*}
    M_{ik}^{(r)} = \frac{M_{ik}}{\sum_{k'}M_{ik'}}, M_{jk}^{(c)} = \frac{M_{jk}}{\sum_{j'}M_{j'k}}
\end{align*}
The spread probability matrix is then calculated as
\begin{align*}
    P_{spread}(i,j) = \sum_kM_{ik}^{(r)}M_{jk}^{(c)}
\end{align*}
Equivalently, in matrix form,
\begin{align*}
    P_{spread} = M^{(r)}(M^{(c)})^\top
\end{align*}
Here, $M^{(r)}$ is the row-normalized matrix representing the destination distribution of travelers from each county, and $M^{(c)}$ is the column-normalized matrix representing the source distribution of travelers arriving at each county.

%




\subsection{Simulated Outbreak}
\label{sec:simulation}
Given the small proportion of cases that have been observed and sampled in real-world settings, we also use simulated data for which the complete disease transmission process is known. We employ an existing large-scale simulation framework that efficiently models the spread of an infectious disease across the entire United States~\cite{kong2025episcale}. In the simulation, we parametrize the disease with a basic reproduction rate, initialize infections randomly in airport regions, and run the simulation for 90 simulation days. The simulation provides exact information on when transmission events occur and who infects whom, which serves as the ground-truth transmission tree. We use 4 different $R_0=[1.25,1.3,1.4,1.5]$ to generate 40 different transmission trees, aiming to test the robustness of our proposed approach facing various scales of infectious diseases. We then use these transmission trees as inputs to an existing genome sequence simulation, $\pi$Buss, generating sequences of length 1701 with a clock rate of 1.8E-3~\cite{bielejec2014pi}. 

\subsection{Sampling With Bias}
\label{sec:bias}
Based on the ``Simulation Ground-Truth'' obtained from the simulation, we further insert synthetic data collection bias, aiming to simulate demographically biased case sampling. This setup allows us to compare the biased sample data to the simulation ground truth~\cite{zufle2024leveraging}. In this paper, we use income as the only attribute affecting the reporting probability of an individual. We obtained the demographics data of each CBG from the US Census Bureau, which splits the income into 16 categories. For each category, given the range of income $[I_{min},I_{max})$, with $I_{mean} = (I_{min}+I_{max})/2$, we define the reporting probability as 
\begin{align*}
    P(\text{report}|I_{mean}) = p_{min} + \frac{p_{max} - p_{min}}{1 + \text{exp}(-k(\text{ln}(I_{mean}+1)-m)}
\end{align*}
with $p_{min} = 0.1$, $p_{max} = 0.9$, $k = 1.2$, and $m=3000$.
We then calculate the average reporting rate for each CBG using the weighted average among all 16 income categories based on the demographic data.

\begin{figure*}
    \centering
    \vspace{-0.3cm}\includegraphics[width=1\linewidth,height=9cm]{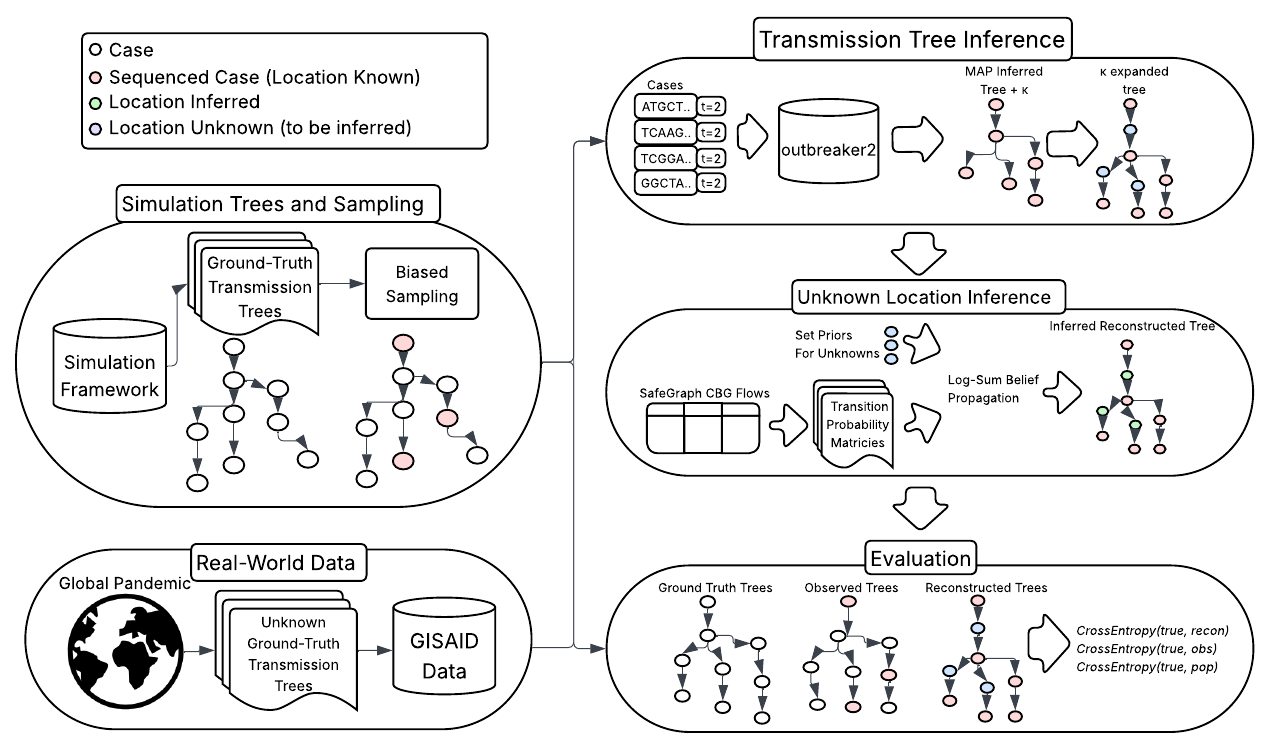}
    \caption{Location inference and bias mitigation workflow: \textbf{Left}: Data generation (simulated) and data preparation (real). \textbf{Right:} Inference of unobserved nodes in the transmission tree using \texttt{outbreaker2} (top); Inference of locations of unobserved cases using belief propagation (middle); Evaluation comparing spatial distributions of inferred and ground truth cases (bottom).  }\vspace{-0.3cm}
    \label{fig:workflow}
\end{figure*}

\subsection{Real-World Sequence Data}\label{sec:gisaid}

SARS-CoV-2 genome sequences and associated metadata were obtained from the GISAID EpiCoV database. Access to GISAID data is governed by the GISAID Database Access Agreement. Because the agreement restricts redistribution of GISAID data, we do not publicly release the downloaded sequences or metadata used in this study. Instead, we provide the GISAID accession identifiers for all genomes included in our analysis (Supplementary Table S1). Researchers with authorized GISAID access can re-download the corresponding records and reproduce our dataset. We gratefully acknowledge all data contributors, including the originating and submitting laboratories responsible for generating and sharing these data through GISAID.

Our dataset consists of 1,437 SARS-CoV-2 genomes collected in Georgia between January 2021 and early 2023. The majority of samples were collected during 2022 (n = 1,076), spanning multiple waves of the Omicron variant.

The spatial distribution of sequencing effort is highly uneven. Bibb County (Macon) contributes approximately 41\% of all genomes (589 of 1,437 sequences), despite representing only about 1.4\% of Georgia's population (approximately 157,566 of 11.3 million residents)  \cite{census_bibb_2024, census_ga_2024}.
This concentration is largely attributable to sequencing efforts conducted by the LSUHS Emerging Viral Threat Laboratory. Another possibility is the lack of county-level labels conducted by other laboratories as part of Georgia's 70k total sequences in GISAID. 

Since the first reported COVID-19 case in Washington State on Jan. 21, 2020, the New York Times tracked cases of COVID-19 in real time as they were identified after testing \cite{nytimes_covid19}. We treat the NYT county-level case counts as our closest available approximation to ground truth, though we note that this dataset itself reflects testing availability and public health reporting practices rather than a representation of the true infection prevalence. In our real-world experiment, we use this distribution as a proxy for actual case geography.

\section{Methodology}
\label{sec:methodology}

                In this section, we describe our system architecture as depicted in Figure~\ref{fig:workflow}. As input, our pipeline takes case samples (including location, time, and genomic sequence) from either (a) simulated data with injected bias as described in Section~\ref{sec:bias}, or (b) real-world sequence data from sequence databases as described in Section~\ref{sec:gisaid}.                
                As detailed in Section~\ref{sec:transmissiontree}, we use \texttt{outbreaker2} to infer a maximum a posteriori (MAP) transmission tree and estimate the number of unsampled intermediate cases ($\kappa$) per branch. For branches with $\kappa > 1$, we insert $\kappa-1$ nodes with Gamma-distributed infection times. We then construct a CTMC rate matrix from Advan's Pattern's CBG mobility level flows \cite{advan}, and run belief propagation on the reconstructed tree to infer location distributions of all unknown cases as detailed in Section~\ref{sec:belief}. We then evaluate our reconstructed geographic distribution against the true outbreak distribution using cross-entropy, comparing against the observed distribution and a population proportional baseline as detailed in Section~\ref{sec:eval_metrics}. 

\subsection{Inference of Transmission Tree}\label{sec:transmissiontree}

Using an existing large-scale simulation framework \cite{kong2025episcale}, we generate ground-truth transmission trees with varying $R_0$. For each tree, we sample with bias as described in Section~\ref{sec:bias}. These sampled cases mimic the observed cases in real-world settings, where sequence and origin are known. We then pass the case sequences and dates into \texttt{outbreaker2} and extract a MAP tree with highest posterior probability after burn-in. 


After we receive an inferred tree from \texttt{outbreaker2}, each  edge $(v_i, v_j)$ has an associated kappa value $\kappa_{ij}$ which represents the number of intermediate generations between observed cases $i$ and $j$. For $\kappa>1$, we insert $\kappa-1$ unobserved cases. However, choosing where to insert the cases should not be a deterministic process, as transmission takes uneven amounts of time dependent on a multitude of factors. Hence, we add case(s) drawn from a Gamma distribution to remain consistent with the \texttt{outbreaker2} likelihood computational structure. For each intermediate case between two observed cases $n_{ij}$, we draw $n_{ij} + 1$ gaps from:

\vspace{-.2cm}

\begin{equation}
    g_k \sim \text{Gamma}(\alpha, \beta), k=1,...,n_{ij} +1
\end{equation}

where $\alpha=4, \beta=2$. We normalize after sampling so that all intermediate times lie within ($t_i, t_j)$. 

\vspace{-.2cm}

\subsection{Location Inference for Unobserved Cases}\label{sec:belief}
Given our augmented transmission tree from \texttt{outbreaker2}, we formally define the problem setup for CBG-level inference of unknown case origin. 

We model CBG-to-CBG transitions using a continuous-time Markov chain (CTMC) with rate matrix $\mathbf{Q} \in \mathbb{R}^{K \times K}$, where $Q_{ab}$ represents the instantaneous rate of transition from CBG $a$ to CBG $b$ for $a \neq b$, and $Q_{aa} = -\sum_{b \neq a} Q_{ab}$.
The rate matrix $\mathbf{Q}$ is derived from observed mobility flows between CBGs. For each directed mobility flow $(a,b)$, we calculate the weight as: 

\begin{equation}
    w_{ab} = \log [(1+ \sum_{res} \operatorname{src\_prob}(a) \cdot \operatorname{des\_prob}(b)])
\end{equation}

where $\text{src\_prob}(a)$ and $\text{des\_prob}(b)$ are the probabilities that an individual originates from CBG $a$ and travels to CBG $b$. We log-scale the resulting matrix to reduce high-traffic CBGs from dominating the rate matrix.

The transition probability over a continuous time interval $\tau$ is given by the matrix exponential:

\begin{equation}
\mathbf{P}(\tau) = e^{\mathbf{Q}\tau}
\end{equation}

As stated in section \ref{subsec:ctmc}, and unlike our previous implementation \citep{tucker2025spatial}, we do not constrain our tree with constant branch lengths. Branch lengths in our augmented tree are derived from infection time between parent and child nodes. 

\subsubsection{Belief Propagation}

Let $T=(V,E)$ be a transmission tree with nodes $V=[{1,...N}]$ where $N$ represents the number of cases with directed edges $E$ representing transmission events. The tree consists of $V_{obs} \subseteq V$ which consists of sequenced cases, of which we know their location. Furthermore, we define $V_{unobs}\subseteq V$ for which we infer missing locations, and $V_{obs} \cup V_{unobs} = V$. Moreover, let $C=[c_1,c_2,...,c_K]$ where K denotes the set of CBGs. Each node $v\in V$ is defined as $v=(t_v, \mathbf{s_v})$ where $t_v\in\mathbf{R^+}$ represents an infection time, and $\mathbf{s_v}$ represents the state vector initialized as: 

\begin{equation}
s_v(c_k) = \begin{cases}
1 & \text{if } v \in V_{\text{obs}} \text{ and county}(v) = c_k \\
0 & \text{if } v \in V_{\text{obs}} \text{ and county}(v) \neq c_k \\
\frac{1}{K} & \text{if } v \in V_{\text{unobs}}
\end{cases}
\end{equation}

where $s_v(c_k)$ denotes the $k$-th component of $\mathbf{s}_v$ corresponding to county $c_k$. For observed nodes, $\mathbf{s}_v$ is a one-hot vector encoding the known county. For unobserved nodes, $\mathbf{s}_v$ is initialized with a uniform prior over all counties.

It should be noted that priors are a pivotal design choice; poorly chosen priors could propagate bias while well-informed priors can reduce it. To explore how priors affect the inferred CBGs of unknown cases, we analyze the same trees under two other prior settings. First, we select a prior that reflects the biased sample distribution, and serves as a baseline for how much BP corrects sampling bias. For each $v\in V_{unobs}$

\begin{equation}
    s_v(c_k) = \frac{{(u\in V_{obs}:  \text{county}(u)=c_k})}{|V_{obs}|}
\end{equation}

Within public health decision making, this represents a prior that assumes there is no bias within sampling, and the sampled distribution represents the true distribution perfectly. We also use a mobility prior from Advan's Mobility Flow \cite{advan}, initialized as the normalized marginal destination frequency of each CBG mobility flow:
\vspace{-.2cm}

\begin{equation}
    s_v(c_k) = \sum_i^K p_{des}(c_k | c_i)
\end{equation}

We use sum-product belief propagation on the augmented tree to produce marginal distributions of CBGs $s_v$ for all nodes. Because we have an acyclic tree structure, we compute the exact marginals within one upward pass and one downward pass. 

\textbf{Upward Pass}
Messages propagate from leaf to root. For each node $j$ with parent $i$, the message sent upward $\boldsymbol{\mu}_{j\rightarrow i}$ is the element-wise product:

\begin{equation}
\mathbf{m}_j = \mathbf{s}_j \odot
\prod_{k \in \mathrm{ch}(j)} \boldsymbol{\mu}_{k \to j}
\end{equation}

\begin{equation}
\boldsymbol{\mu}_{j \to i} = P(\tau_{ij}) \, \mathbf{m}_j
\end{equation}

where $\mathbf{s}_j$ represents the current belief state of node $j$, and $\boldsymbol{\mu}_{k\rightarrow j}$, $k\in ch(j)$ represents the message(s) sent from all children of node $j$. Our resulting $\mathbf{m}_j$ is multiplied by the transition probability matrix from the rate matrix $Q$ at some branch length time $\tau_{ij}$.

\textbf{Downward Pass}
Once messages propagate upward, we compute a downward pass. Messages propagate from the root towards the leaves. Let $\boldsymbol{\nu}_{i\rightarrow j}$ represent the message sent from node $i$ to its child $j$. We combine the downward message received by $i$, excluding the contribution from child $j$. 

\begin{equation}
    \boldsymbol{\nu}_{i\rightarrow j} = P(\tau_{ij})^\top [\nu_{p(i)\rightarrow i} \odot (\mathbf{m}_i \oslash \boldsymbol{\mu}_{j\rightarrow i})]
\end{equation}

Where $\mathbf{m}_i \oslash \boldsymbol{\mu}_{j\rightarrow i}$ represents the element-wise division of the message from child $j$. The resulting belief is propagated using the transition matrix $P(\tau_{ij})$. Our posterior belief at node $j$ is an accumulation of upward and downward evidence:

\begin{equation}
    \mathbf{b}_j \propto \mathbf{m}_j \odot \boldsymbol{\nu}_{p(j)\rightarrow j}
\end{equation}

To obtain our final CBG-level prediction for a node $j$, we select the highest probability state from each node's belief distribution:
\vspace{-.4cm}

\begin{equation}
    CBG_{j} = \arg \max_c({b}_j(c))
\end{equation}

\subsection{Evaluation Metrics and Baseline}\label{sec:eval_metrics}

We evaluate all three distributions, observed, reconstructed, and true, at a common timepoint $T^*$, defined as the infection time of the 50th sequenced case. We fix $N=50$ as a representative early-outbreak cutoff, as comparable to or earlier than thresholds used in literature that define the onset of exponential outbreak growth \cite{kochanczyk2020super}. This mirrors early outbreak settings where public health officials have limited sequences in hand, and must characterize current geographical spread. By restricting all distributions to cases with $t\leq T^*$, each distribution describes the same window of outbreak. We report Cross-Entropy (CE) on the three distributions in comparison to the true distribution. Cross Entropy measures how well our reconstructed distribution matches the true outbreak distributions.
Our ground truth transmission trees for simulation data are the sampled and unsampled cases up until $t=T*$, while we use the NYT dataset \cite{nytimes_covid19} as proxy for our true distribution in our real-world experiment. 

We compare our results to a baseline of a population distribution, normalized to sum to one across CBGs, which represents public health decision making where case prevalence is assumed to be directly linked to population. For the simulation and real-world experiments, we use the 2020 U.S. Census Bureau to obtain population for each CBG and county respectively \cite{uscensus_gazetteer}. It should be noted that there is a significant discrepancy between the CBGs present in our mobility flows and the U.S. Census Bureau data. Some CBGs are not present within the Census data ($17,463/25,305$ available). Hence, when calculating cross entropy, we only include CBGs present within both our mobility flow and the U.S. Census Bureau dataset. While this inherently lowers cross-entropy across all distributions, it ensures a fair comparison between our proposed approaches.

\section{Experimental Evaluation}
\label{sec:experiment}
To illustrate that our pipeline can de-bias early infectious disease reporting data, and provide a more accurate assessment of early outbreak spread, we test on 40 simulated trees with varying $R_0$. We create our observed cases through biased sampling, and use a down-sampled tree by restricting each tree to its first 50 observed cases. We pass each down-sampled tree into \texttt{outbreaker2}, and expand the generated transmission tree. After obtaining our marginals for each node, we evaluate using cross-entropy. We also provide our results on real-world data based in Georgia, USA, from GISAID. 

All the source code for the experiments presented in the following are found in GitHub at \url{https://github.com/OwenTucker/Phylogeographic-Inference-Model}. The simulated and synthetically biased case data is found in this GitHub as well. Due to data redistribution restrictions imposed by GISAID, we are unable to publicly share the raw sequence datasets directly. To ensure full experimental reproducibility, we provide the complete list of EPI\_ISL IDs that uniquely identify our sequenced cases. Interested readers can retrieve the exact sequences and associated metadata by querying these identifiers directly through the GISAID Database.

\vspace{-0.25cm}
\subsection{Results on Simulated Data}

We generated 40 ground-truth transmission trees with an equal split across $R_0=[1.25, 1.3, 1.4, 1.5]$. Full simulated trees range from 409 to over 100,000 cases, with the 50-day cutoff capturing diverse outbreak lengths between 17 and 60 days. Each tree was sampled at a $20\%$ sampling rate, including bias as described in Section~\ref{sec:bias}. \texttt{Outbreaker2} is used to infer a transmission tree from these 50 sequences, $\kappa$-based interpolation reconstructs unsampled intermediate cases, and belief propagation parameterized by mobility flow produces a reconstructed geographic distribution over the CBG state space.  


We provide our results on 40 simulated trees with sampling bias. We aggregate over $R_0$, and report: i) Cross entropy of the true and observed distributions, ii) Cross entropy of the true and reconstruction distributions, iii) Cross entropy of the true and population-density distribution baseline. We also report using three different priors as defined in Section~\ref{sec:methodology}. We test over three priors as an ablation study to observe the effects of different initialization, where each prior choice represents different interpretation of outbreak characterization. This isolates the effect of latent-node initialization from the effect of mobility-informed propagation itself, which is held constant. For instance, a uniform prior represents an equal likelihood for each CBG regardless of population. An empirical prior represents using our observed cases as evidence of where unknown cases lie. Lastly, a mobility prior assumes destination frequencies are correlated with outbreak spread. Our results are shown in Table~\ref{tab:ce_r0_prior_pop}. 

\begin{table}[t]
\centering
\caption{Cross-entropy results by $R_0$ and prior ($n=10$ trees per cell). Bold indicates the values which beat $CE_{pop}$ baseline; all cross-entropy values are evaluated against the true transmission tree distribution.}\vspace{-0.4cm}
\label{tab:ce_r0_prior_pop}
\begin{tabular}{lcccccc}
\toprule
 & & \multicolumn{3}{c}{$\text{CE}_\text{rc}$ by Prior} & & \\
\cmidrule(lr){3-5}
$R_0$ & $\text{CE}_\text{ob}$ & Uniform & Mobility & Empirical & $\text{CE}_\text{pop}$ \\
\midrule
1.25 & 6.9521 & \textbf{3.8989} & \textbf{4.1501} & \textbf{5.2060} & 6.0865 \\
1.30 & 7.7515 & \textbf{4.3104} & \textbf{4.7127} & 7.1587          & 6.1585 \\
1.40 & 6.7459 & \textbf{4.2125} & \textbf{4.4796} & \textbf{5.3459} & 7.5398 \\
1.50 & 6.2755 & \textbf{3.8277} & \textbf{4.0324} & \textbf{5.7527} & 7.0913 \\
\bottomrule
\end{tabular}
\vspace{-0.8cm}
\end{table}

The reconstructed transmission tree consistently beats the population baseline in terms of cross-entropy, and hence better represents the true distribution of cases. In an early stage of an outbreak, population density is not representative of the true distribution of cases. However, comparing only cross entropy of the observed cases to the population baseline, we would make the opposite conclusion, as for $R_0 <1.3$, our population baseline beats the cross entropy of observed cases. It is clear that using our mobility-based approach, we are able to create a reconstructed transmission tree that reduces observable bias and better characterizes the true nature of the outbreak. Our results bridge the gap between the two main failure modes for early-outbreak resource allocation: i) if public health officials act solely upon the sequenced cases, they risk over-allocating to counties with testing bias and under-allocate to counties where transmission is actually occurring, and ii) if public health officials use population as a proxy for missing cases, resources are spread proportionally to population regardless of where the outbreak actually is, missing the fact that early outbreaks are geographically condensed and not yet population proportional. Our results outperform both of these standards in early-outbreak scenarios. 

As expected, a prior based on observed cases performs poorly compared to the other approaches, as this only reinforces the sampling bias present. However, even with a poor initialization, it still improves upon the observed cross entropy. What is worth noting is that the uniform prior consistently performs the best of all three. This suggests that for early-outbreak scenarios, the strongest signal is the transmission tree's structure, and a uniform prior allows propagation of the initial beliefs to move freely. Mobility's primary contribution is through this propagation, and not through prior initialization. 

As well as measuring agreement between reconstructed and true spatial distributions, we also evaluate whether the reconstructed tree correctly identifies locations in which transmission is most concentrated. This distinction is important, as two distributions can have similar error while differing substantially in how they prioritize high-risk locations. Using both the population baseline and our reconstructed trees, we define high-risk locations as fraction of infections within the true transmission tree. We compare the rankings produced to the true transmission tree. Across all 40 simulated trees, our reconstructed tree achieved a mean Spearman correlation of ($0.4735 \pm 0.1679$), compared with ($-0.0297 \pm 0.1144$) for the population baseline. Hence, our framework not only reduces bias in observed cases, but properly identifies high-risk locations.

\vspace{-0.4cm}
\subsection{Experiments on Real World Data}
To illustrate our method upon real-world genomic surveillance data, we draw upon GISAID data. The entire dataset is vast, and spans multiple years and across different variants of COVID-19. To create plausible transmission trees, we split the genomic data based on clade, or major lineages. We selected clades with total cases $n>30$ as inputs to outbreaker. 

A significant challenge with using GISAID data is that location attached to sequenced cases is generally lab-based, rather than originating from the location of the case itself. Hence, to avoid trees with overwhelming lab-based bias, we discard trees with $>90\%$ case prevalence within a single county. Cases that are not assigned to any clade are dropped from our analysis. After selecting the sub-trees that fit our criteria, we process each in \texttt{outbreaker2}. We run belief propagation on the augmented tree to obtain marginals for the unobserved cases added from \texttt{outbreaker2}. 

We provide our results on six real-world COVID-19 inferred transmission trees. It should be noted that as real-world data is recorded at a county level, our inferred case locations are also on a county level. Our results are shown in Table~\ref{tab:gisaid_ce_results}.

\begin{table}[t]
\centering
\caption{Cross-entropy results on real-world GISAID trees by lineage. $N_\text{obs}$ = sequenced cases, $N_\text{tot}$ = total tree size (observed + latent), $C$ = number of CBGs spanned, $L$ = infection length (days). Cross-entropy is calculated using the NYT dataset as the ground truth proxy.}
\label{tab:gisaid_ce_results}
\begin{tabular}{lcccccccc}
\toprule
Lineage & $N_\text{obs}$ & $N_\text{tot}$ & $C$ & $L$ & $\text{CE}_\text{ob}$ & $\text{CE}_\text{rc}$ & $\text{CE}_\text{pop}$ \\
\midrule
BA.1  & 381 & 1003 & 18 & 163 & 3.947 & 3.544 & \textbf{2.563} \\
BA.5  & 306 & 1519 & 34 & 228 & 3.800 & 3.693 & \textbf{3.606} \\
BA.2  & 240 & 1098 & 29 & 260 & 3.311 & \textbf{3.395} & 3.692 \\
BA.4  &  90 &  470 & 15 & 209 & 2.757 & 2.641 & \textbf{2.153} \\
BQ.1  &  46 &  288 &  3 & 174 & 2.002 & 1.100 & \textbf{0.781} \\
BF.10 &  31 &   92 &  5 & 113 & 2.710 & 2.217 & \textbf{1.918} \\
\bottomrule
\end{tabular}
\end{table}

Across all six lineages, the reconstructed distribution consistently achieves lower cross-entropy than the observed distribution, ranging from 0.107 (BA.5) to 0.902 (BQ.1). This demonstrates that mobility-based belief propagation reliably reduces geographic bias introduced by lab-based location sequencing, regardless of lineage or outbreak size. The smaller lineages (BQ.1, BF.10) show the largest relative improvements, consistent with our simulation findings that bias correction adds the most value when our observed sample is small and localized. 

However, the population baseline achieves lower cross-entropy than our reconstructed distribution in a majority of lineages. This is attributed to the fundamental differences between our method's design and the nature of the available COVID-19 genomic data. Our simulation experiments illustrate that mobility-based reconstruction outperforms population-based inference specifically in early-outbreak scenarios. However, the real-world lineages span from 113 to 260 days, during which tens of thousands of cases accumulated across Georgia counties. At this scale, the NYT ground truth distribution is large enough that demographic priors are inherently strong predictors. This is structurally distinct from early outbreak settings, as COVID-19's multi-wave epidemic does not represent the early-stage outbreak where phylogeographic correction is most actionable. 

\vspace{-.25cm}


\section{Conclusion}
\label{sec:conclusion}
Spatial data on infectious diseases is frequently compromised by sampling bias across subpopulations and location bias, either stemming from cases being recorded at testing laboratories rather than transmission sites, or inherent limitations in testing availability. In this paper, we introduced a novel framework that leverages large-scale human mobility data alongside phylogenetic tools to mitigate these biases. Our approach builds on the intuition that mobility networks can bridge observation gaps: flow patterns between regions enable the system to infer the true spatial distribution of underrepresented cases. Formally, our technical pipeline inputs sequenced case data into the \texttt{outbreaker2} framework to estimate missing transmission links, and applies belief propagation to infer the locations of these latent cases. We validated our method using both simulated and empirical data. In controlled simulations with synthetic bias, our approach significantly reduced the divergence between biased observations and the known ground truth, including against a population-based baseline. The cross entropy improvements reported translate directly to more accurate geographic probability estimates over CBGs at a time where such estimates are most actionable. At a low number of observed sequences, the true outbreak has already spread beyond the biased surveillance sample but has not reached a population-proportional distribution, a window in which geographic characterization informs containment rather than mitigation. By reducing cross entropy by $\approx40\%$ relative to observed data while outperforming population-based baselines in simulated trees, our method provides a more accurate characterization of transmission using genomic surveillance data.   
When applied to real-world GISAID sequence data, our framework successfully corrected bias stemming from local sequencing facilities. Our model shifted the inferred spatial distribution closer to the true historical case counts.
We acknowledge the limitations of Advan's Patterns data as shown by Coston et al. \cite{coston2021leveraging}. Moreover, future work should incorporate more sources of bias besides solely income. 
In summary, anchoring disease surveillance within human mobility networks provides a robust mechanism to correct missing or mislabeled spatial data, ultimately offering public health authorities a more accurate foundation for targeted interventions.
\vspace{-.3cm}


\section{Acknowledgments}
This research is supported by the Australian Commonwealth Scientific and Industrial Research Organisation (CSIRO) and the United States National Science Foundation (NSF) under Grant Nos. 2302968, 2302969, and 2302970 (titled ``Collaborative Research: NSF-CSIRO: HCC: Small: Understanding Bias in AI Models for the Prediction of Infectious Disease Spread''), with additional independent support from the NSF under Grant No. 2109647. We gratefully acknowledge the Authors, Originating and Submitting Laboratories of the sequences from the GISAID EpiCoV\texttrademark{} Database on which this research is based. \vspace{-0.3cm}


\bibliographystyle{ACM-Reference-Format}
\bibliography{refs/main}

\end{document}